\documentclass[pra,showpacs,floatfix,twocolumn,superscriptaddress]{revtex4}
\usepackage{graphicx,amsmath,amssymb}
\usepackage{epsf,color}
\usepackage{floatflt,latexsym}

\begin{document}
\title{Strong violations of Bell-type inequalities for Werner-like states}  
\author{Christoph~F.~Wildfeuer}
\email{wildfeuer@phys.lsu.edu}
\author{Jonathan~P.~Dowling}
\affiliation{Hearne Institute for Theoretical Physics, Department of Physics
  and Astronomy, Louisiana State University,
 Baton Rouge, Louisiana 70803, USA}
\begin{abstract}
We investigate the violation of Bell-type inequalities for two-qubit Werner-like
states parametrized by the positive parameter $0\le p\le 1$. We use an
unbalanced homodyne detection scheme to obtain the quantum mechanical
probabilities. A violation of the Bell-Wigner and Janssens inequalities is obtained for a large range of
the parameter $p$. The range given by these inequalities is greater than
the one given by the Clauser-Horne inequality. The range in which a violation is attained actually coincides
with the range where the Werner-like states are known to be nonseparable, i.e., for
$p>1/3$. However, the improvement over the Clauser-Horne inequality is
achieved at the price of restricting the class of possible local hidden
variable theories.
\end{abstract}
\maketitle
\section{Introduction}\label{Intro}
\par Since the seminal paper of Einstein, Podolsky, and Rosen \cite{EPR} on
the completeness and physical reality of quantum mechanics, a great deal of research has been devoted to clarify the relation between
nonseparable quantum states and nonlocal correlations. This relation, as it
turns out, is far from trivial. For an overview see, e.g., Ref.~\cite{Genovese}. 
\par If a quantum state is nonseparable, i.e, cannot be factored in some product
of its subsystems, it is called
entangled. Although a state may be entangled, it need not necessarily manifest
nonlocal correlations. It was Bell that suggested an operational criterion to test
the predictions of local hidden variable (LHV) theories against quantum
mechanical predictions. The violation of a Bell inequality by a specific
quantum state is then an indication that the state is able to exhibit nonlocal correlations. Almost all current
experimental tests of nonlocality are based on the Clauser-Horne (CH)
inequality \cite{Clauser-Horne}, the Clauser-Horne-Shimony-Holt
(CHSH) inequality \cite{CHSH}, and a few on the Wigner inequality \cite{Wigner}.   
\par It is known that for any entangled {\em pure} state of any number
of quantum systems one may violate a generalized Bell inequality \cite{Gisin,
  Popescu}. An extension of this statement for mixed entangled states has not been
found. Furthermore, Werner \cite{Werner} provided in 1989 an example of nonseparable mixed states that
do not violate the CHSH inequality. He then constructed
  a LHV model which replicated the properties of his mixed quantum states,
  which have since become known as Werner states. This demonstrated that one
  cannot always violate Bell inequalities with any mixed entangled
  state. Two-qubit Werner states are parametrized by the positive parameter $0\le p\le
  1$ and can be expressed as follows:
\begin{eqnarray}
  \rho^\mathrm{W}_p=p|\Psi\rangle\langle \Psi|+\frac{(1-p)}{4}\mathbb{I}\,.\label{Werner1}
\end{eqnarray}
The Werner states are a mixture of an entangled pure state $|\Psi\rangle$ with noise.  
Werner showed that these states admit a LHV model for
  projective measurements for $1/3< p\le 1/2$ and violate the CHSH inequality for $p> 1/\sqrt{2}$. 
It is known that these states are separable if and only if $p\le 1/3$
\cite{Werner}. A number of authors have further investigated the nonlocal properties of Werner states
\cite{Gisin0, Barett, Acin, Popescu2, Teufel}. Recently,
Ac\'{i}n {\em et al.} \cite{Acin} have extended the range for which a LHV
model for Werner states can be constructed.
It is not known whether Werner states admit a LHV model
for projective measurements in the region $0.66< p \le 1/\sqrt{2}$, or if
there is a Bell inequality that may be violated, which refers
to the gap in Fig.~\ref{werner_overview}.
Inspired by this open problem it is interesting to investigate the amount of violation obtained
for various Bell-type inequalities with Werner states or states of similar
structure than Werner states. It appears that the amount of nonlocality obtained in an experiment
depends on the specific Bell-type inequality tested and the measurement carried
out.
\par A very powerful geometrical interpretation of Bell-type inequalities was
introduced by Pitowsky in 1986, where they are referred to as correlation
polytopes \cite{Pitowsky86}. He showed that the probabilities in a
Bell-type inequality can be considered to be a vector in a convex
polytope. One can think of the points contained within this polytope as each
representing a set of measurement probabilities satisfying a Bell-type
inequality, which is represented by the bounding planes of the polytope
itself. This enables one to use tools from convex geometry to construct
other Bell-type inequalities relevant for other quantum tests of nonlocality
\cite{mathematica}. Recently Janssens {\em et al.} \cite{Fuzzy} generated in
this way Bell-type inequalities in the correlation polytope for six
joint probabilities which may be used to test local realism in physical
experiments. The structure of these inequalities is very similar to the Bell-Wigner
inequality. 
\par The article is organized as follows: We start our investigation with an inequality from the well-known
Clauser-Horne polytope. We then investigate
the violation of an inequality from the lesser well-known Bell-Wigner polytope which contains three joint probabilities. Finally we present an analysis of Janssens' Bell-type inequalities
in the correlation polytope for six joint probabilities. The single and joint
probabilities for the Bell tests are calculated explicitly for an unbalanced
homodyne detection scheme, where the two detectors are assumed to be spacelike
separated. We consider the two-qubit Werner
state from Eq.~(\ref{Werner1}) for all Bell tests presented. Note that
$\mathbb{I}$ is the identity operator in the
two-qubit space. For the pure entangled state $|\Psi\rangle$ we consider the one-photon entangled state
\begin{eqnarray}\label{onephoton} 
|\Psi\rangle=\frac{1}{\sqrt{2}}(|1\rangle_a|0\rangle_b-|0\rangle_a |1\rangle_b)\,,
\end{eqnarray}
where the labels $a$ and $b$ refer to the two path-entangled modes accessible
to Alice and Bob, respectively.  
\par There has been a long debate about the actual nature of
entanglement for the one-photon entangled state because it contains only one particle. A few authors
argued that at least two particles are required to make a proper entangled state. Some arguments
can be found in Refs.~\cite{vanEnk1,vanEnk2} and citations therein. Van
Enk has since pointed out that the two field modes themselves
should be considered entangled rather than the individual photons --- an
approach in which the dispute appears to be settled \cite{vanEnk2}. 
\par Although this state can now clearly be called entangled until very
recently it has been questioned whether or not a single photon entangled state
exhibit nonlocal correlations. Since the very thorough investigation by Dunningham and
Vedral \cite{Dunningham}, there can be no doubt anymore that single particles
in a delocalized state can exhibit nonlocal correlations. The authors
furthermore point out that ``we must not view nonlocality as pertaining
to particles themselves, but see it instead as a property of quantum fields
whose significance is, therefore, more fundamental than that of particles.''  
Some recent experiments are also in agreement of
the nonlocal nature of the one-photon entangled state discussed in this
article \cite{Lvovsky}.
\par Note that many authors use the above notation
to refer to spin up and down, i.e., $|1\rangle=|\uparrow\rangle$ and
$|0\rangle=|\downarrow\rangle$, which refers to a different physical
system and hence leads to other correlation functions, i.e., probability
distributions. The
results obtained in Refs.~\cite{Gisin0, Barett, Acin, Popescu2, Teufel} are
obtained for the spin-$1/2$ singlet and the usual traceless spin projection
operators. Therefore they do not need to coincide
with our results obtained for the one-photon entangled state. Note also the
projection operators given in Eqs.~(\ref{povm1}), and (\ref{povm2}) are not
traceless. For the sake of clarity we emphasize that we refer to {\em
  Werner-like} states in our article. We leave the term {\em Werner states} for the original approach
in which the maximally entangled pure two-qubit state is considered to be the
spin-1/2 singlet {\em and} the measurement operators are the spin projectors.
\section{Unbalanced homodyne detection scheme}\label{correlationmeasurement}
\par
\begin{figure}[htb]
\includegraphics[scale=0.4]{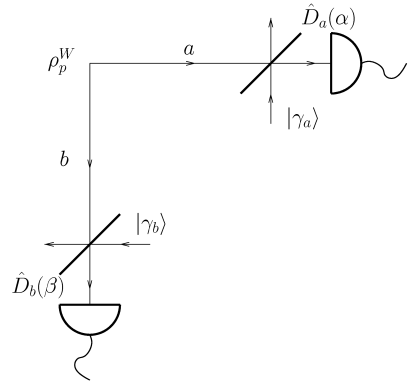}\caption{Unbalanced homodyne detection scheme for a Bell experiment with
  Werner-like states. Here $\rho^\mathrm{W}_p=p|\Psi\rangle\langle
  \Psi|+\frac{(1-p)}{4}\mathbb{I}$ with $|\Psi\rangle=\frac{1}{\sqrt{2}}(|1\rangle_a|0\rangle_b-|0\rangle_a|1\rangle_b)$
  and $a$ and $b$ label the modes. Alice and Bob mix mode matched local
  oscillators of amplitude $\gamma_a$ and $\gamma_b$, respectively, with their
  signal modes, where $\alpha=\gamma_a\sqrt{1-T}$ and
  $\beta=\gamma_b\sqrt{1-T}$ are the arguments of the displacement operator $\hat{D}$
  defined in the text. The symbol $T$ denotes the transmittivity of the beam
  splitters.\label{homodyne}}
\end{figure}

 We calculate the quantum mechanical probabilities for our Bell test for an
unbalanced homodyne detection scheme as displayed in Fig.~\ref{homodyne}.
State reconstruction with an unbalanced homodyne detection scheme has been
investigated in detail by Wallentowitz and Vogel
\cite{Wallentowitz}. The authors showed that the $Q$ function can be
measured directly in an unbalanced homodyne detection scheme with a perfect detector by recording the probability of zero
counts as a function of the complex amplitude of the local oscillator.
The photon detectors are considered to be simple on-off detectors, i.e.,
  they cannot resolve the photon number. The local projective measurements are described by $\hat{Q}(\alpha)+\hat{P}(\alpha)=\hat{{\bf 1}}$, with 
\begin{eqnarray}
\hat{Q}(\alpha)&=&\hat{D}(\alpha)|0\rangle\langle 0|\hat{D}^\dagger(\alpha)\,,\label{povm1}\\
\hat{P}(\alpha)&=&\hat{D}(\alpha)\sum_{n=1}^\infty|n\rangle\langle
n|\hat{D}^\dagger(\alpha)\,.\label{povm2}
\end{eqnarray}
The displacement operator is defined by
\[\hat{D}(\alpha)=\mathrm{exp}(-\frac{1}{2}|\alpha|^2)\mathrm{exp}(\alpha\,\hat{a}^\dagger)\mathrm{exp}(-\alpha^\ast
\hat{a}),\] \cite{Gerry}.
The expectation value of $\hat{Q}(\alpha)$ tells us the probability that no photons are present, depending on the phase and amplitude of the local
oscillator.  The expectation value of $\hat{P}(\alpha)$ gives the
probability of counting one or more photons. We obtain a binary result by
assigning a one to a detector click and a zero otherwise. The corresponding
measurement operators for a correlated measurement of the displaced vacuum can
be written as $\hat{Q}_a(\alpha)\otimes\hat{Q}_b(\beta)$. The joint probability for the Werner-like state is calculated from
 \begin{eqnarray}\label{joint}
   Q_{ab}(\alpha,\beta)=\mathrm{Tr}\left(|\alpha\rangle_a\langle\alpha|\otimes|\beta\rangle_b\langle\beta|\,\rho^\mathrm{W}_p\right)\,,
\end{eqnarray} 
and we obtain
 \begin{eqnarray}\label{jointresult}
   \lefteqn{Q_{ab}(\alpha,\beta)=\frac{p}{2}e^{-|\alpha|^2-|\beta|^2}(|\alpha-\beta|^2)}\nonumber\\
                &&{}+\frac{1-p}{4}e^{-|\alpha|^2-|\beta|^2}(1+|\alpha|^2+|\beta|^2+|\alpha|^2|\beta|^2)\,.         
\end{eqnarray}
The single-count probability for Alice's measurement is given by
 \begin{eqnarray}\label{single}
   Q_{a}(\alpha)=\mathrm{Tr}\left(|\alpha\rangle_a\langle\alpha|\otimes\hat{{\bf
   1}}_b\,\rho^\mathrm{W}_p\right)=\frac{1}{2}e^{-|\alpha|^2}(|\alpha|^2+1)\,,
\end{eqnarray} 
and the corresponding probability for Bob's measurement of $Q_b(\beta)$ is 
\begin{eqnarray}\label{single2}
   Q_{b}(\beta)=\mathrm{Tr}\left(\hat{{\bf
   1}}_a\otimes|\beta\rangle_b\langle\beta|\,\rho^\mathrm{W}_p\right)=\frac{1}{2}e^{-|\beta|^2}(|\beta|^2+1)\,.
\end{eqnarray}
Alice and Bob obtain the same probability distribution for their independent
measurements. Note also that the joint probabilities in
Eq.~(\ref{jointresult}) are symmetric in the two arguments, i.e., $Q(\alpha,\beta)=Q(\beta,\alpha)$. We
may therefore drop the subindices on the $Q$ functions for what follows. We can assume that the
first argument for the joint probabilities $Q(\alpha,\beta)$ is Alice's local
oscillator (LO) setting and the second argument belongs to Bob's LO. The
single-count probabilities can essentially be obtained from a run of the
experiment at either Alice's or Bob's laboratory. 
With the above single and joint probabilities, we can test if any of the Bell-type inequalities are violated.
\section{Violation of Clauser-Horne and Wigner inequalities}
\par
\begin{figure}[htb]
\includegraphics[scale=0.4]{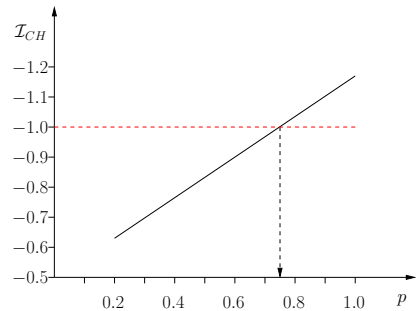}
\caption{(Color online) Violation of the CH inequality $-1\le \mathcal{I}_{CH}\le 0$ with $\mathcal{I}_{CH}$ defined in
  Eq.~(\ref{CH}) as a function of the mixing parameter $p$. The horizontal dashed line marks
  the classical constraint. We obtain a
  violation for $p>0.75$.\label{violation_CH}}
\end{figure}

 We investigate the violation of the well-known CH inequality
\cite{Clauser-Horne}, given by $-1\le\mathcal{I}_{CH}\le 0$, where $\mathcal{I}_{CH}$ is defined as follows: 
\begin{eqnarray}
\mathcal{I}_{CH}&=&Q(\alpha,\beta)-Q(\alpha,\beta')+Q(\alpha',\beta)+Q(\alpha',\beta')\nonumber\\
& &{}-Q(\alpha')-Q(\beta).\label{CH}
\end{eqnarray}
A minimization method for the parameters $\alpha,\alpha',\beta$, and $\beta'$ results
in a violation as a function of $p$ as displayed in Fig.~\ref{violation_CH}.
We attain a violation for any $p>0.75$. If we consider the Werner state in the
spin-1/2 basis and decide to make a spin correlation measurement instead, we
can violate the CH inequality for any $p>1/\sqrt{2}\approx 0.71$, which is
well known \cite{Werner}. 
\par The Werner-like state in the Fock basis together with a
measurement that projects onto coherent states, as given by Eq.~(\ref{povm1}), obviously leads to a different result
for the range of violation than the Werner state in the spin basis and a spin correlation measurement. 
Note also that the operators given in Eqs.~(\ref{povm1}), and (\ref{povm2}) are not
traceless. Therefore the known bounds for traceless operators as derived in Ref.~\cite{Acin} cannot be directly applied.   
\par Next we consider one of the inequalities from the Bell-Wigner polytope
\cite{Pitowsky} given by $W_1\le 1$, where 
\begin{eqnarray}
  W_1&=&Q(\alpha)+Q(\beta)+Q(\gamma)\nonumber\\
  & &{}-Q(\alpha,\beta)-Q(\alpha,\gamma)-Q(\beta,\gamma).\label{Wignerinequality}
\end{eqnarray}
Note that original Bells' inequality \cite{Bell} cannot be applied to
noisy entangled states such as the CH and CHSH can. This is due to the fact that Bell's
approach works only for perfect anticorrelated pure states. This assumption is
no longer fulfilled for mixed entangled states where the correlations may not
be perfect anymore. In a recent paper by Pitowsky it is shown that Bell's
original approach can be extended to mixed entangled states if the
states are not too noisy \cite{Pitowsky2}.
Pitowsky's extension of Bell's inequality
to mixed entangled states also works for the Bell inequalities we are going to
consider in this and the next section. Pitowsky also considers two limiting cases which make further assumptions about the
distribution of the local hidden variables. The inequalities we are
considering are valid for the
assumption of zero average symmetry breaking \cite{Pitowsky2}. We note that the
extra assumptions on the distribution of the local hidden variables restrict
our approach to a specific class of local hidden variable models. However, the
fair sampling assumption, which is commonly assumed in all experiments, already rules out a large class of local hidden
variable models.    
\par
\begin{figure}[htb]
\includegraphics[scale=0.4]{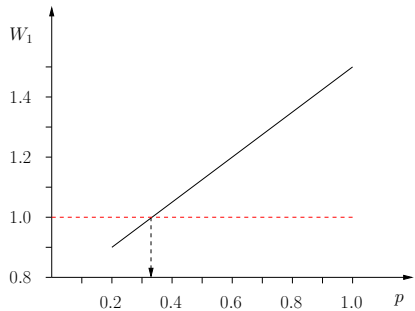}
\caption{(Color online) Violation of the Bell-Wigner inequality $W_1\le 1$ with $W_1$ defined
  in Eq.~(\ref{Wignerinequality}) as a function of the mixing parameter
  $p$. The horizontal dashed line marks the classical constraint. We obtain a violation for $p>1/3$.\label{Wigner_violation}}
\end{figure}

 A maximization method for the complex parameters $\alpha$, $\beta$, and
$\gamma$ results in a violation as a function of $p$ as displayed in
Fig.~\ref{Wigner_violation}.
From Fig.~\ref{Wigner_violation} we see that the range $p$ for which the value
$W_1$ violates the classical constraint is very large. A
further analysis shows that one may numerically approach the
separability-nonseparability border at $p=1/3$ with a high precision. We
see that the above test of the Bell-Wigner inequality shows a violation for the
two-qubit Werner-like state for the entire nonseparability range $p>1/3$. The
Bell-Wigner inequality is obviously a stronger inequality than the
Clauser-Horne inequality,
in the sense that it shows a larger range of violation for the Werner-like
state.
\section{New Bell tests} 
\par
\begin{figure}[htb]
\includegraphics[scale=0.4]{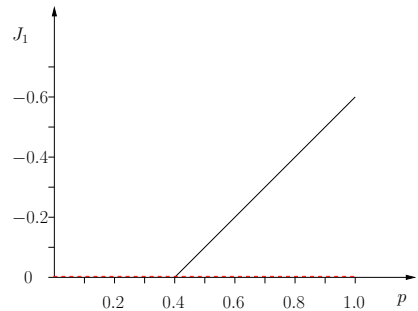}
\caption{(Color online) Violation of the Bell-type inequality $0\le J_1$, where $J_1$ is defined
  in Eq.~(\ref{six_11}), as a function
  of $p$. The horizontal dashed line marks the classical constraint. We obtain a violation for $p>0.4$.\label{ch1}}
\end{figure}

 We finally consider Bell-type inequalities which, to our knowledge, are
not currently used in Bell tests. These Bell-type inequalities belong to the correlation polytope for
six joint probabilities, which have been constructed by Janssens {\em et al.}
\cite{Fuzzy}. We display them first in terms of classical single and joint
probabilities, where the probability of a single random event $A_i$ is defined
by $p_i=P(A_i)$ and the joint probability of a pair of random events $A_i$ and
$A_j$ is denoted by $p_{ij}=P(A_i\cap A_j)$: 
 \begin{equation}
  0\le p_i+p_j+p_{ij}-p_{ik}-p_{i\ell}-p_{j\ell}-p_{jk}+p_{k\ell}\,,\label{six_1}
\end{equation}
 \begin{equation}p_i+p_j+p_k+p_\ell-p_{ij}-p_{ik}-p_{i\ell}-p_{jk}-p_{j\ell}-p_{k\ell}\le 1\,,\label{six_2}
\end{equation}
\begin{equation} 
2p_i+2p_j+2p_k+2p_\ell-p_{ij}-p_{ik}-p_{i\ell}-p_{jk}-p_{j\ell}-p_{k\ell}\le 3\,,\label{six_3}
\end{equation}
\begin{equation}
0\le p_i-p_{ij}-p_{ik}-p_{i\ell}+p_{jk}+p_{j\ell}+p_{k\ell}\,,\label{six_4}
\end{equation}
\begin{equation} 
p_i+p_j+p_k-2p_\ell-p_{ij}-p_{ik}+p_{i\ell}-p_{jk}+p_{j\ell}+p_{k\ell}\le 1\,,\label{six_5}
   \end{equation}
for any different $i,j,k,\ell$.
We investigate the amount of violation of
the inequalities in Eqs.~(\ref{six_1})--(\ref{six_5}) for the simple on-off
detection scheme, with the probabilities given by Eqs.~(\ref{jointresult}),
(\ref{single}), and (\ref{single2}). The probabilities for the inequality (\ref{six_1}) are then replaced by
\begin{eqnarray}
  J_1=Q(\alpha)+Q(\beta)+Q(\alpha,\beta)-Q(\alpha,\gamma)-Q(\alpha,\delta)\nonumber\\
  -Q(\beta,\delta)-Q(\beta,\gamma)+Q(\gamma,\delta)\label{six_11},
\end{eqnarray}
so that the inequality is expressed by $0\le J_1$.
We display the results for the minimization procedure in Fig.~\ref{ch1}.
We obtain a violation for $p>0.4$ which is linearly increasing with
$p$ until it reaches the maximum for $p=1$. For the violation of the pure
one-photon entangled state compare with Ref.~\cite{Wildfeuer} also. 
\par
\begin{figure}[htb]
\includegraphics[scale=0.4]{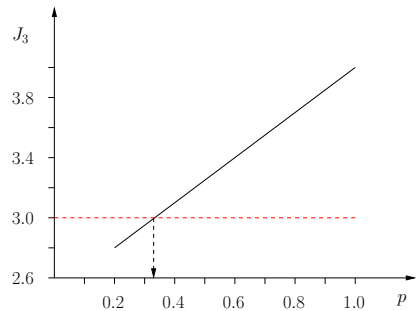}
\caption{(Color online) Violation of the Bell-type inequality $J_3\le 3$, where $J_3$ is
  defined in Eq.~(\ref{six_31}), as a function
  of $p$. The horizontal dashed line marks the classical constraint. We obtain a violation for $p>1/3$.\label{ch3}}
\end{figure}

 As another example of the violations obtained in this correlation
polytope, we investigate the inequality (\ref{six_3}). This can be
rewritten in terms of the local oscillator amplitudes as well, where we
introduce the quantity $J_3$ by
\begin{eqnarray}
J_3&=&2Q(\alpha)+2Q(\beta)+2Q(\gamma)+2Q(\delta)-Q(\alpha,\beta)\nonumber\\
& & {}\hspace{-3mm}{-Q(\alpha,\gamma)-Q(\alpha,\delta)-Q(\beta,\gamma)-Q(\beta,\delta)-Q(\gamma,\delta)}\label{six_31},\nonumber\\
\end{eqnarray}
to write the inequality in the form $J_3\le 3$.
In Fig.~\ref{ch3} we show again the violation as a function of $p$.
The range of violation is exactly the same as for the Bell-Wigner inequality, where we obtain a violation for any $p>1/3$.
\section{Conclusion}
\par
\begin{figure}[htb]
\includegraphics[scale=0.7]{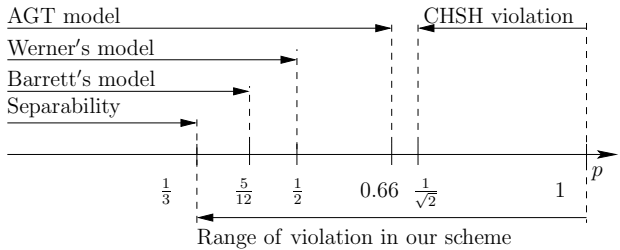}
\caption{Overview of the current status on the violation of Bell-type
  inequalities in different models by two-qubit Werner states
  $\rho^\mathrm{W}_p$ as a function of $p$. The models refer to the following
  publications: Barrett \cite{Barett}, Werner \cite{Werner}, Ac\'{i}n, Gisin, and
  Toner (AGT) \cite{Acin}.\label{werner_overview}}
\end{figure}

 We conclude from our results that
Bell-type inequalities exist, which are violated for the entire range in which
Werner-like states are nonseparable, i.e., entangled. In the unbalanced homodyne
detection scheme, we showed that the Werner-like states
violate one of the well established Bell-Wigner inequalities for the entire
range $p>1/3$. We also presented Bell-type
inequalities in the correlation polytope of six joint probabilities that have
not been investigated so far in the context of
experiments on local realism. Some of these inequalities also give a violation
for the full range where the Werner-like states are entangled. 
We may therefore add to the overview given in Ac\'{i}n {\em et al.'s} publication
\cite{Acin} our result and obtain Fig.~\ref{werner_overview}.
 Our result also shows that Bell-type inequalities other than
the CH or CHSH are less sensitive when noise is added to the pure entangled state
$|\Psi\rangle$. Several authors, see, e.g., Refs.~\cite{Collins, Acin} and the
references
therein, have investigated the robustness of Bell tests against noise. The inequalities from the
Bell-Wigner and Janssens correlation polytope are obviously advantageous for
Bell tests under the influence of noise. It should be noted, however, that for our
specific measurement scheme, this advantage does not 
seem to be related to the dimensionality of the correlation polytope,
since the Bell-Wigner inequality with three joint probabilities shows the
same violation as the Janssens inequality with six joint probabilities. In
fact, this noise-resistant feature seems much more closely tied to the
method used to construct the joint probabilities. The Bell-Wigner inequalities
allow Alice, for example, to perform a
measurement with the
LO settings $\alpha$ and $\beta$ and Bob with $\beta$ and $\gamma$. We see
that one setting, here $\beta$, is assigned once to Alice's laboratory and in
another run of the experiment to Bob's laboratory. Pitowsky
showed that a local hidden variable model can be constructed for this combination of events, so that the violation of
the Bell-Wigner inequality is a true indication for the violation of the
locality assumption. However our approach is only valid for a restricted class
of local hidden variable models in which zero average symmetry breaking is assumed \cite{Pitowsky2}. Note that Janssens' inequalities similarly combine the
measurements which
Alice and Bob may perform, merely taking more combinations into account.
This is different from the
structure in the CH inequality. Here Alice accesses the LO settings $\alpha$
and $\alpha'$ while Bob can perform a measurement with $\beta$ and
$\beta'$. Alice and Bob always have distinct LO settings in their
laboratories. We point out that our
example demonstrates a way in which other Bell-type inequalities can be
valuable.  
\begin{acknowledgments}
C.F.W. and J.P.D. acknowledge the ARO and the DTO for support. We thank H. Lee,
S.J. Olson, W. Plick, M.M. Wilde, K. Jacobs, N. Sauer, R.~Kretschmer,
M.P. Seevinck, J. Bergou, and B. He for very helpful comments.
\end{acknowledgments}

\end{document}